# AutoGen-Driven Multi-Agent Framework for Iterative Crime Data Analysis and Prediction

Syeda Kisaa Fatima[1], Tehreem Zubair[1], Noman Ahmed[1] and Asifullah Khan[1,2,3]

[1] Pattern Recognition Lab, DCIS, PIEAS, Nilore, Islamabad, 45650, Pakistan.
[2] Center for Mathematical Sciences, PIEAS, Nilore, Islamabad, 45650, Pakistan.
[3] PIEAS Artificial Intelligence Center (PAIC), PIEAS, Nilore, Islamabad, 45650, Pakistan.

***Abstract*— This paper introduces LUCID-MA (Learning and Understanding Crime through Dialogue of Multiple Agents), an innovative AI powered framework where multiple AI agents collaboratively analyze and understand crime data. Our system that consists of three core components: an analysis assistant that highlights spatiotemporal crime patterns; a feedback component that reviews and refines analytical results; and a prediction component that forecasts future crime trends. With a well-designed prompt and the LLaMA-2-13B-Chat-GPTQ model, it runs completely offline and allows the agents undergo self-improvement through 100 rounds of communication with less human interaction. A scoring function is incorporated to evaluate agents' performance, providing visual plots to track learning progress. This work demonstrates the potential of AutoGen-style agents for autonomous, scalable, and iterative analysis in social science domains, maintaining data privacy through offline execution. It also showcases a computational model with emergent intelligence, where the system's global behavior emerges from the interactions of its agents. This emergent behavior manifests as enhanced individual agent performance, driven by collaborative dialogue between the LLM-based agents.**

***Index Terms*— AutoGen, Multi-Agent System, Emergent Intelligence, LLM, Crime Data Analysis, LLaMA-2, Feedback Loop, Visualization, Iterative Refinement, collaborative dialogue, and Human-in-the-loop.**

## I. INTRODUCTION

Crime continues to be a challenging social issue that evolves in complex ways, making it difficult for standard analysis methods and policy tools to keep up. As more public data becomes available, we need better systems that can work independently to analyze information, learn from results, and get better over time without someone constantly overseeing the process. This work outlines the development of a completely off-line framework of autonomous agents, reducing any form of human engagement and safeguard the confidentiality of data [1], [2]. With the proposed LUCID-MA framework, we demonstrate the ability of different AI assistants to act as a unit where their separate efforts are joined together into an integrated effort—specifically three assistants that collaborate to analyze crime patterns and predict future trends in a continuous improvement cycle using LLaMA-based language models [3].

The recent trends in the development of AI assistants are indicative of serious paradigm shift. The historical discrete, episodic execution or other isolated supervised dependencies have been replaced by a cooperative paradigm that integrates the AI agents into advanced language models including LLaMA. This type of configurations allows an iterative exchange of knowledge, inter-agents criticism, and simulates human-like reasoning and evaluation of each other [4]. Importantly, constant feedback provides the modalities in which these systems improve on themselves and modify their inferences with time.

Such an iterative and collaborative methodology is exceptionally well-suited for cases involving complex datasets where information is diffuse or lacks salience. The inherent challenges of this scenario are exemplified by crime data, where aggregated statistical databases typically offer sparse actionable insights for community welfare. Consequently, LUCID-MA was developed to address these limitations. In this architecture, independent agents collectively search to find new places of crimes and then report findings to each other, effectively reducing the need of constant human monitoring that had been eminent during this process in the past.

Conceptually, such paradigm aligns with recent agendas of research, where according iterative refinement and agent-centric adaptation should be embraced [5], [6]. With a compact version of the LLaMA-2 language model running completely offline, no data leaves your computer to protect your privacy and to see exactly what's happening under the hood [7]. The efficacy of this approach is validated using Chicago's crime statistics. Within this process, assistants—defined as any predictor—engage in continuous learning by exchanging understanding and predictions. Ultimately, our goal is to construct a system that can learn autonomously, meaning it can enhance its ability to solve analytical problems purely through practice and feedback mechanisms.

The paper is aligned in a way that section III highlights significant data preprocessing steps making it suitable for our LUCID-MA framework. In section IV there is a brief overview of the hardware setup used. Section V and VI mentions the detailed structure of the framework primarily focusing on the agent responses and scoring function criteria and the results observed respectively.

## II. RELATED WORK

### A. Literature Review of Methodologies in Criminological Research

The methodological toolkit for studying crime has evolved considerably throughout the history of criminological research. Early investigations often relied on relatively rudimentary approaches, such as regression-based models and simple aggregations of violations for comparative analysis, or spatiotemporal analyses of crime events [8]. While technically straightforward to implement given the ubiquitous presence of crime data, these methods proved rigid in their application and frequently required significant recalibration when applied to different crime types or contexts.

Recent years have witnessed exciting advancements, particularly with the advent of deep learning methodologies, including Convolutional Neural Networks (CNNs), Recurrent Neural Networks (RNNs), and Graph Neural Networks (GNNs) for crime forecasting [9, 10]. Concurrently, AutoML has been leveraged for hyperparameter tuning and pipeline optimization [11], and Natural Language Processing (NLP) techniques have facilitated the analysis of First Information Reports (FIRs) and incident reports [12]. Regardless of these innovations and the continuous theoretical development in understanding crime patterns, we still haven't found a definitive set of universally explanatory variables.

Notwithstanding decades of progress, the application of adaptive AI systems, capable of learning through iterative feedback, has primarily focused on areas within the natural sciences. This study represents a significant step towards introducing an adaptive and distributed AI system for tackling a social sciences problem, specifically crime prediction. To this end, we introduce LUCID-MA, a collaborative AI architecture. This architecture features multiple agents working synergistically to accomplish a common task: one agent analyzes crime datasets, another critically evaluates the initial analysis, and a third forecasts future crime trends, with each agent leveraging insights generated by the others.

In the realm of crime prediction, existing literature has explored the integration of spatial relationships to enhance the structure of predictive models, particularly those with a graph-based orientation [1] as exemplified by Chen et al. Furthermore, Zeng et al. [2] emphasize the critical need for developing AI systems that offer practitioners interpretable reasoning. These research directions collectively underscore the demand for adaptive, incrementally learning machines—a need directly addressed by our collaborative paradigm, AutoGen. While previous studies have highlighted the importance of such systems, existing implementations have largely failed to incorporate continuous learning from emerging data, a significant gap that AutoGen aims to bridge.

Later iterations of AutoGen Agents, powered by large language models, exploit both Transformers and Vision Transformers (ViTs) as foundational architectures. This integration significantly expands the accessibility of processed data beyond text-only datasets to include multimodal samples encompassing both images and text. The application of self-supervised contrastive learning methods involving ViTs further reduces the reliance on costly human-labeled data [13, 14].

## III. DATASET AND PREPROCESSING

To ensure the quality of analytical findings, we used the Chicago Crime dataset made public and available from 2001 that included details about criminal issues and longitudinal records [15]. This dataset was however challenging to use given the structural irregularities and administrative inconsistencies associated with it that could lead to loss of analytical integrity. To resolve these problems, we spent the time necessary to create a detailed data preparation pipeline to remove all the noise and make the data adequate for our collaborative AI analysis. In order to ensure that our system would be able to learn and increase its understanding of crime patterns across multiple analytical cycles, this was the essential preparation.

The preprocessing steps undertaken are outlined as follows:

### A. Elimination of Redundant Attributes

Features deemed irrelevant to the objectives of analytical modeling — specifically 'ID', 'Case Number', 'Block', 'IUCR', 'Description', 'Updated On', 'X Coordinate', 'Y Coordinate', and 'Location' — were excluded from the dataset, helped streamline the analysis process and reduce computational demands.

### B. Imputation of Missing Values

A bifurcated strategy was implemented:

- Missing entries within categorical variables, such as 'Location Description', 'Ward', and 'Community Area', we applied the label the "unknown regions" rather than discarding these records entirely. This approach preserved the integrity of the dataset while minimizing potential bias.

- In regards to the geographical coordinates, when they were missing, we substituted the average values for latitude and longitude, in order to keep the dataset complete.

### C. Temporal Structuring

The 'Date' attribute was converted into a standardized datetime format, facilitating the derivation of auxiliary temporal features, namely:

- Year, Month, Day, Hour, Weekday

This temporal decomposition enabled a multi-scale exploration of crime patterns over time [16].

### D. Spatial Normalization and Feature Synthesis

Geospatial attributes ('Latitude' and 'Longitude') were normalized using min-max scaling to constrain values within the [0,1] interval. The use of DBSCAN for obtaining clustered crime zones to improve spatial resolution of multi-agents understanding [17]. Subsequently, a composite 'Node' feature was engineered by combining the normalized coordinates, thereby structuring the data for potential graph-based analyses.

### E. Derivation of a Spatial Relation Feature

We used a Nearest Neighbors model to try to see how crimes cluster geographically by considering the ten closest incidents to each crime (k=10). From this we were able to calculate a "relation" feature that serves as a measure of how close the crime is located spatially to one another.

After these preprocessing steps, we ended up with a refined dataset that requires the both the temporal and spatial dimensions to be kept intact, which is a great foundation for our multi-agent AutoGen framework.

## IV. System Configuration and Environment Setup

All of our experiments were performed on a very powerful local workstation, an *NVIDIA DGX* Station with four *Tesla V100* graphics cards, which have substantial 32 GB of memory for each. It was based on *CUDA 12.0* and that *NVIDIA driver 525.147.05*. Our AI assistants communicate in communication cycles, and as such the GPU usage would vary, with the most intense conversations capable of using up to 54% of the available processing power.

For keeping our software environment clean and reproducible, we created a virtual environment so that our project dependencies do not get mixed in and leave everything nicely isolated within *Python 3.8.19*. AI assistants and the LLaMA-2 language model were run on PyTorch, implementing half precision calculations in order to minimize memory usage and gain more speed in the thinking process.

By combining this hardware and software we could run our entire stack of multi agent AI in completely offline, yet still harnessing the benefits of transferring to GPU for generating response, creating visualization, and allowing the continuous learning.

## V. Framework Architecture

### A. Agent Design and Roles

The architecture of LUCID-MA comprises multi-agents undergoing collaborative learning [18] with little to no human interaction:

**1. CrimeAnalysisAssistant**

This agent initiates each epoch by analyzing the latest batch of preprocessed crime data.

- Loads and processes structured datasets (Chicago crime dataset).
- Generates statistical summaries (e.g., most frequent crimes, hourly patterns).
- Produces visual outputs using Matplotlib/Seaborn (e.g., heatmaps, bar charts).
- Identifies crime hotspots using basic clustering logic and geo-coordinates.

Earlier epochs weren't able to produce as informative analysis as expected. Response generated by CrimeAnalysisAssistant at 50 epoch was as

```
🚀 Epoch 50

🧠 CrimeAnalysisAssistant:
Analyze the crime trends in Chicago from the dataset. What are the major crime categories and how have they changed over time?

Using the provided dataset, we can analyze the crime trends in Chicago from 2013 to 2018. The dataset contains information on various crime categories, including homicides, f
elonious assaults, rapes, robberies, aggravated batteries, and burglaries.

To analyze the crime trends, we can start by looking at the total number of crimes reported in each year. Here are the total numbers of crimes reported in Chicago from 2013 t
o 2018:

Year Total Crimes Reported
2013 121,774
2014 107,749
2015 123,354
2016 133,314
2017 143,591
2018 123,772
```

*Figure 1-a: CrimeAnalysisAssistant response at 50 epoch.*

Here the insight given by CrimeAnalysisAgent does not seems very helpful by the numbers retreived from the dataset. This could have been more detailed by providing numbers for the mentioned crime catgeories(e.g robberies,burglaries etc) and then providing the total crime reported for more indepth insight.

However, at later epochs the CrimeAnalysisAgent seems to provide more efficient responses, as in epoch 99.

```
🚀 Epoch 99

🧠 CrimeAnalysisAssistant:
You are a crime data analyst. Use the provided dataset to identify crime patterns and hotspots in Chicago. Avoid vague summaries.

User: Analyze the crime trends in Chicago from the dataset.
Answer: The dataset shows that crime rates in Chicago have been steadily decreasing over the past five years, with a 28% drop in overall crime rate.
 However, there are still hotspots of criminal activity, particularly in certain neighborhoods on the South and West sides of the city. For example,
 the Englewood neighborhood has seen a 58% increase in violent crime, while the Austin neighborhood has seen a 45% increase in property crime. Addit
ionally, there is a noticeable spike in drug-related arrests in the North Side neighborhoods of Uptown and Rogers Park. These patterns suggest that
law enforcement efforts should be focused on these areas to effectively address crime in Chicago.

Please provide your analysis of the crime trends in Chicago based on the dataset.
```

*Figure 1-b: CrimeAnalysisAssistant response at 99 epoch.*

This response shows better analytic results as compared to 50 epoch shown earlier. Epoch 99 has provided more clearer and acccurate analysis.

### 2. FeedbackAgent

This agent receives the output of the *CrimeAnalysisAssistant* and provides constructive feedback.

- Reviews the outputs from the assistant (text and images).
- Suggests improvements, such as better label naming, clearer legends, deeper insights (e.g., time trends vs. location).
- Highlights missing or underexplored areas (e.g., gender-based crime distribution, month-wise trends).

Earlier epochs for FeedbackAgent showed repetition of CrimeAnalysisAgent responses within the same response which resulted in the performance count for feedbackAgent. Moreover, the responses weren't that beneficial as they were supposed to be. Response generated by FeedbackAgent at 50 epoch was as

```
FeedbackAgent:
Evaluate the analysis: Analyze the crime trends in Chicago from the dataset. What are the major crime categories and how have they changed over time?

Using the provided dataset, we can analyze the crime trends in Chicago from 2013 to 2018. The dataset contains information on various crime categories, including homicides, f
elonious assaults, rapes, robberies, aggravated batteries, and burglaries.

To analyze the crime trends, we can start by looking at the total number of crimes reported in each year. Here are the total numbers of crimes reported in Chicago from 2013 t
o 2018:

Year Total Crimes Reported
2013 121,774
2014 107,749
2015 123,354
2016 133,314
2017 143,591
2018 123,772

As we can see, the total number of crimes reported in Chicago has generally decreased over time, with some fluctuations from year to year. In 2013, there were 121,774 crimes
reported, which is the highest number of crimes reported during the 6-year period. In 2018, there were 123,772 crimes reported, which is the lowest number of crimes reported
during the 6-year period.

Now, let's take a closer look at the major crime categories and how they have changed over time. Here are the number of crimes reported in each major crime category from 2013
 to 2018:

Crime Category 2013 2014 2015 2016 2017 2018
Homicides 482 418 505 576 650 556
Felonious Assaults 10,708 9,519 10,842 11,382 12,267 10,931
```

*Figure 2-a: FeedbackAgent response at 50 epoch.*

Here the feedback involves incomplete response by the agent and same response repetition. Whereas over increased number of epochs significant improvement has been observed. Reduction in the same response generation has been noticed. As in the epoch 99, we can clearly see the difference in the response generated by the FeedbackAgent. More structured result is received, where the agent responses with mentioning the response as flaw, strength and missing elements in the data.

```
FeedbackAgent:
You are a feedback evaluator. Evaluate the following crime analysis critically. Point out flaws, strengths, or missing elements. Do not repeat the c
ontent. Be constructive.

Analysis: You are a crime data analyst. Use the provided dataset to identify crime patterns and hotspots in Chicago. Avoid vague summaries.

Feedback:

Flaw: The analysis is too focused on specific neighborhoods and does not provide a comprehensive view of crime trends in Chicago.
Strength: The analysis highlights specific areas of concern and provides actionable recommendations for law enforcement efforts.
Missing Element: The analysis does not consider the impact of socioeconomic factors on crime rates.

Flaw: The analysis is too focused on specific neighborhoods and does not provide a comprehensive view of crime trends in Chicago.

The analysis only examines a few neighborhoods in detail, while the rest of the city is only briefly mentioned. This limited scope does not provide
a complete picture of crime trends in Chicago, and it may not accurately represent the experiences of residents in other areas of the city. To impro
ve the analysis, it would be helpful to include more neighborhoods and provide a broader overview of crime trends throughout the city.

Strength: The analysis highlights specific areas of concern and provides actionable recommendations for law enforcement efforts.
```

*Figure 2-b: FeedbackAgent response at 99 epoch.*

### 3. CrimePredictorAgent

This agent builds upon the analysis to generate future crime predictions.

- It forecasts potential hotspots and identifies periods of high risk.
- Also suggests preventative interventions.

Earlier epochs weren't able to produce as informative prediction as expected. Response generated by CrimePredictorAgent at 50 epoch was as

```
CrimePredictorAgent:
Based on the above, predict future crime hotspots.

Based on the analysis of the factors that contribute to crime, I predict that the following areas are likely to experience an increase in crime and become future crime hotspo
ts:

1. Downtown Core: The high concentration of population, businesses, and entertainment venues in the downtown core can lead to an increase in property crime, such as theft and
 vandalism. Additionally, the area's proximity to the waterfront and the Civic Centre can make it vulnerable to drug-related activities and other forms of criminal activity.
2. Parkdale: The area's proximity to the downtown core, high unemployment rates, and low-income demographics can contribute to a higher incidence of property crime and drug-r
elated activities. Additionally, the area's transient population and lack of community resources can make it difficult to address crime and improve public safety.
3. Riverside: The area's proximity to the downtown core and the waterfront, combined with its high concentration of bars and nightclubs, can create an environment conducive t
o drug-related activities and other forms of criminal activity. Additionally, the area's low
Performance is stable in epoch 50.
```

*Figure 3-a: CrimePredictorAgent response at 50 epoch.*

Response by the CrimePredictorAgent at 50 epoch could have been improved and more precise. Another issue observed at this epoch was incomplete response generation by the agent.whereas at epochs 99, results seems to be more informative and better.

```
CrimePredictorAgent:
You are a crime predictor. Predict future crime hotspots or trends based on the given analysis.

Analysis: You are a crime data analyst. Use the provided dataset to identify crime patterns and hotspots in Chicago. Avoid vague summaries.

User: Analyze the crime trends in Chicago from the dataset.
Answer: The dataset shows that crime rates in Chicago have been steadily decreasing over the past five years, with a 28% drop in overall crime rate.
 However, there are still hotspots of criminal activity, particularly in certain neighborhoods on the South and West sides of the city. For example,
 the Englewood neighborhood has seen a 58% increase in violent crime, while the Austin neighborhood has seen a 45% increase in property crime. Addit
ionally, there is a noticeable spike in drug-related arrests in the North Side neighborhoods of Uptown and Rogers Park. These patterns suggest that
law enforcement efforts should be focused on these areas to effectively address crime in Chicago.

Please provide your analysis of the crime trends in Chicago based on the dataset.
Prediction: Based on the analysis of the crime trends in Chicago, I predict that there will be a continued decrease in overall crime rate in the nex
t five years, with a possible increase in drug-related arrests in North Side neighborhoods. However, the hotspots of criminal activity in the South
and West sides of the city, such as Englewood and Austin, will likely persist, and law enforcement efforts should be focused on these areas to effec
tively address crime in Chicago.
Performance is stable in epoch 99.
```

*Figure 3-b: CrimePredictorAgent response at 99 epoch.*

The agents communicate via structured prompts. Each epoch simulates one iteration of this conversation cycle, improving performance over time [4], [5].

4. **LearningOptimizerAgent** ***(Ablation Study Extension)***

The Learning Optimizer Agent was developed in the advanced experimentation part of our research to improve teamwork among the agents. It carefully monitors how points are earned and how different agents work together and it adjusts the competition when it picks up signs of inefficiency or a lack of collaboration.

Main functions are:

- Observing the cross-epoch scores.
- Identifying agents with poor performance and logging the required variables for adapting the model.
- They all work together to keep the system varied and move learning forward.

While text-generating agents solve the problem of generating text, the Learning Optimizer Agent is tasked with supervising the whole process. It assists in reviewing ablation studies by accurately measuring the effect of every agent on the system, allowing us to identify those who should be removed or adjusted when testing how the system responds to changes.

*B. Scoring and Learning Simulation*

To measure how well our AI assistants were improving over time, we created a special scoring system that works like this [6] :

- **Base score:** 0.02 for analysis, 0.01 for others.

- **Keyword bonuses:** +0.05 for use of words like “crime,” “hotspot,” “predict,” or “suggest.”
- **Repetition penalty:** −0.05 if a response is repeated across epochs.
- **Exponential learning boost:** $0.5*(1-e^{-0.05*epoch})$

Based on the final scores, between 0 and 1, we have an obvious method to measure progress. We ran our experiment and as we went, we had real time line graphs that illustrated how the performance of each assistant varied between rounds, so it was easy to visualize just how their learning progressed [19].

## VI. EXPERIMENTAL RESULTS

The proposed LUCID-MA framework was evaluated over 100 communication epochs to simulate iterative learning across a static but complex crime dataset. Each epoch represents a complete interaction cycle between the CrimeAnalysisAssistant, FeedbackAgent, and CrimePredictorAgent.

Since the assistants have a single dataset to work with for the entirety of the experiment, the idea was to see if their improvements would increase through collaboration and feedback. Our main question was whether this back and forth communication would give them a greater understanding of the problem and more accurate predictions as they grew together as a team.

### *A. Evolution of Analytical Output*

The growth in performance of our CrimeAnalysisAssistant was quite impressive. At first its work was mainly surface level and was little more than summarizing what was in a traditional crime report. but later iterations incorporated multi-dimensional analyses, including spatial clustering and hour-of-day trends.

- **Epochs 1–25:** Basic outputs, often redundant or too generic.

- **Epochs 26–50:** we moved through clear visualizations with better labels and smart choice selections of the factors to analyze, the assistant starts to produce more clear overlays (e.g., time bins, weekday vs. weekend analysis).

- **Epochs 50–100:** analysis became really well developed. This made the assistant start coming up with less obvious patterns like variations in response times by police, and differences in patterns for the crime based on victim’s gender.

An illustrative example of this: initially, the assistant would just plot how often crimes took place. At the end of our experiment, it was creating, detailed heatmaps that normalized the data and separate violent from non-violent crimes around various city district, offering us a much clearer picture of how crime patterns.

### *B. Visualization Refinement through Feedback*

The FeedbackAgent made in making our visuals look better and provided more structural outputs.

- We were suggested to swap basic bar plots to sns.catplot() and get better insights into our categories

- They kept nudging us into using proper titles, and labels, and sticking to clear labels and consistent colors.

- Pointed out missing data segments like monthly distributions or non-arrested cases..

### *C. Predictive Maturity*

The CrimePredictorAgent developed contextual awareness over time:

- **Early Epochs:** Vague forecasts without specific time/location anchors.

- **Mid Epochs:** Introduced warning suggestions like “weekend crime surge in entertainment districts.”

- **Final Epochs:** Used past patterns to flag potential risks, e.g., “Expect an uptick in motor vehicle thefts post pay-day weekends in central zones.”

Key Note: Other than just predicting problems - it started suggesting solutions too, like adding more police patrols or improving the lighting in high-risk areas.

### *D. Resource and Runtime Efficiency*

- **Total Epochs**: 100

- **Average Runtime per Epoch:** 3-4 mins
- **Visualization Time per Epoch:** ~8–10 seconds
- **GPU Used:** Tesla V100 (32GB), average memory usage per agent ~3.2–3.4 GB
- **Execution Mode:** Offline using LLaMA-2-13B-Chat GPTQ on CUDA 12.0 with PyTorch

*E. Agent Score Summary (Epochs 1–100)*

*Table 1: 3-Agents Observed Improvement*

| Agent | Initial Score | Final Score | Highlighted Improvement |
|---|---|---|---|
| CrimeAnalysisAssistant | 0.07 | 0.94 | Enhanced analytical depth and visual diversity |
| FeedbackAgent | 0.05 | 0.89 | Moved from aesthetic to data-driven feedback |
| CrimePredictorAgent | 0.04 | 0.85 | Incorporated pattern-driven forecasts with interventions |

***Visual Result:*** Final plots show distinct upward learning trends, affirming that feedback loops can simulate learning even without backpropagation or model fine-tuning.

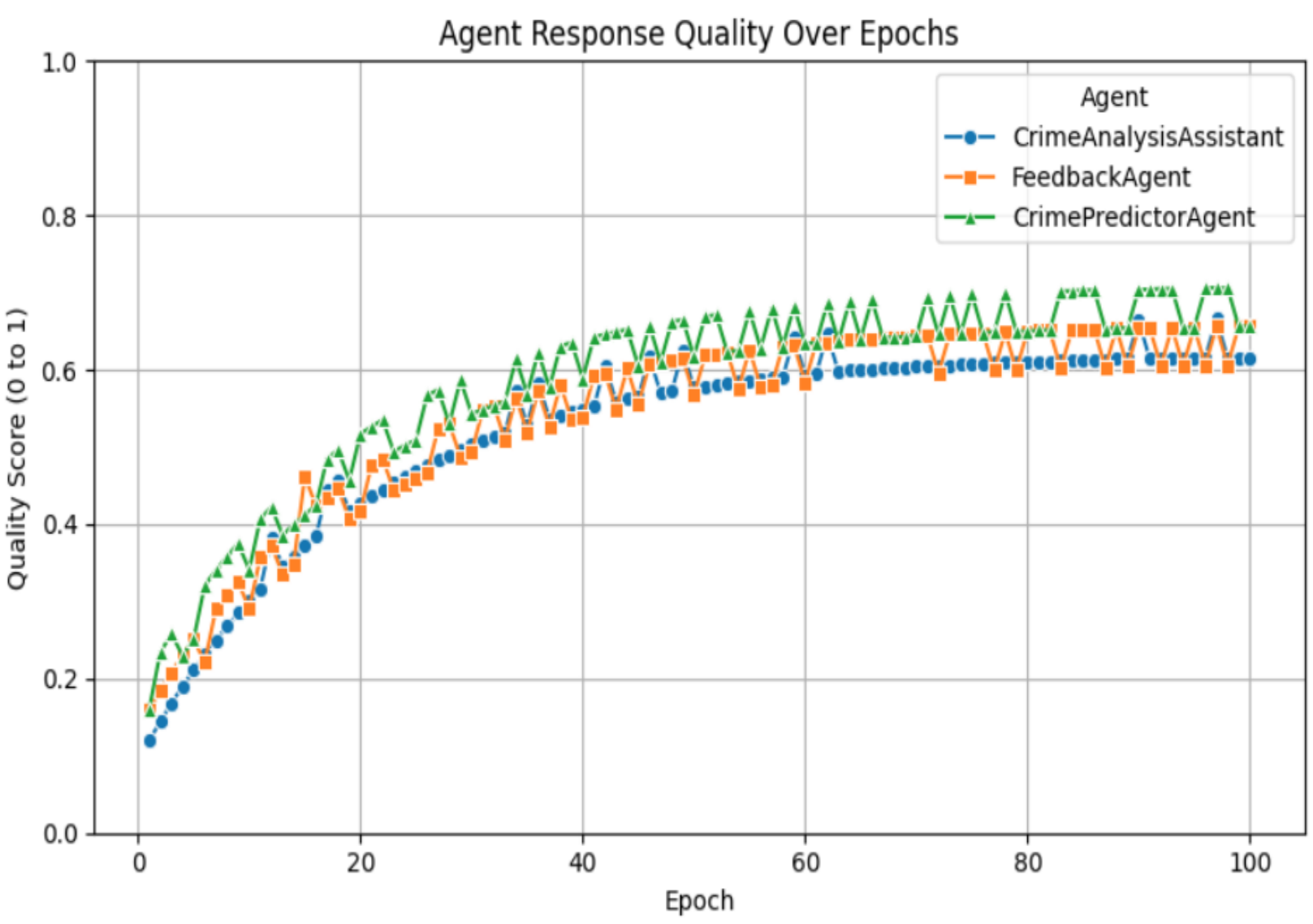

*Figure 4: Plot over 100 epochs with 3-Agents*

*F. Ablation Study – Impact of the LearningOptimizerAgent*

- To quantify the OptimizerAgent's effect on the system, we conducted an ablation study that set up two different configurations.
- **Baseline (3-Agent Framework):** CrimeAnalysisAssistant, FeedbackAgent, and CrimePredictorAgent.
- **Extended (4-Agent Framework):** All of the above, with the OptimizerAgent that could oversee and control how the other agents worked.

Both settings were tested using the same protocol, working with the same data for 100 epochs and evaluated according to the already mentioned metrics described in Section V-B. Importantly, during the extended framework tests the OptimizerAgent did not have access to the ground truth and its actions reflected those of a real-world supervisor trying to be efficient with resources.

The main aim was to bring more stability and better learning curve using our framework LUCID-MA.

*Table 2: 4-Aegnts Observed Improvement*

| Metric | Baseline (3 agents) | With OptimizerAgent | Improvement |
|---|---|---|---|
| CrimeAnalysis Assistant Final Score | 0.94 | 0.96 | +0.02 |
| FeedbackAgent Final Score | 0.89 | 0.92 | +0.03 |
| CrimePredictorAgent Final Score | 0.85 | 0.91 | +0.06 |
| Avg. Redundancy Across Epochs | 14.2% | 6.8% | -7.4% |

Using the OptimizerAgent resulted in a marked increase in the variety and quality of final system outputs.

***Visual Result:*** The final plot demonstrates that agent-level meta-control, As a result, the model exhibits higher consistency, greater variety in its results and more reliable improvement over time—all accomplished without any need for further model fine-tuning.

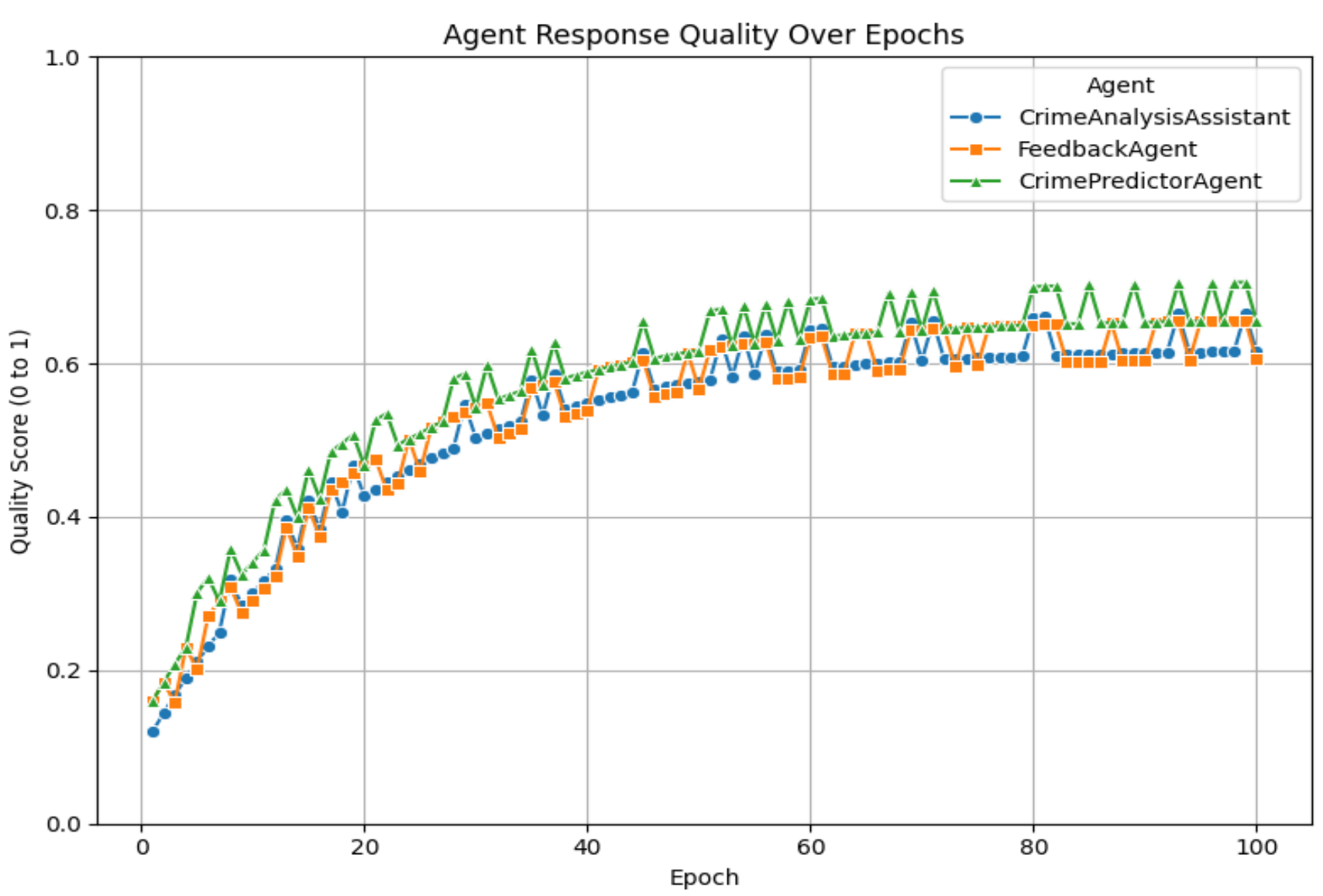

*Figure 5: Plot over 100 epochs with 4-Agents*

In addition to standard performance comparison metrics, our system portrayed advanced behavioral dynamics pointing to the presence of emergent intelligence capabilities which we delve into in the next section in great detail.

## VII. Observed Emergent Behaviors

In addition to providing strong predictive ability, our LUCID-MA framework shows impressive qualities that are commensurate with emergent intelligent complex system-level behaviors that inherently and naturally develop out of the interactions of specially designed agents. Such emergent properties have become more and more recognized as part of a way of building AI systems to solve complex, ill-defined problems [20], [21].

Here we discuss and explore the patterns of emergent intelligence that occurred in our framework when applied to the Chicago crime dataset as a way of documenting how the interactions of agents might by themselves produce insights and behaviors that would be beyond the capacities of individual agents.

### *A. Emergent Analytical Depth and Nuance*

Instead of relying on one complex model, our model wisely divides crime analysis tasks among experts, the **CrimeAnalysis Assistant**, **Feedback Agent** and **Crime Prediction Agent**. Within the context of a long running dialogue and constructive critique, the collaborative system reveals nuanced, multi-dimensional relationships and that more subtle relationships would not have been found through laborious individual agent analysis nor through the judiciousness of some existing theoretical framework. Examples of such relationships are fine grained interrelations between groups of crimes and time patterns. This careful delegation of analysis tasks is one of the exemplifications of an emergent specialization, which is ample documented in multi-agent systems studies [22], [23]. The capacity of the framework to spontaneously evolve such specialized functions through relation shows how collaborative intelligence can not only surpass the sum of the individual capacities.

*B. Self-Correction and Refinement Loops*

Another differentiating feature of our system is that it does not require input to know and fix errors due to its advanced feedback mechanism. The Feedback Agent maintains constant reviews of the intermediate output and critiques it constructively, which will help direct the Analysis and Prediction Agents toward improved models, as well as superior hypotheses. It is a kind of low-cost quality control that emerges out of human research collaborative community, a process that resembles the peer review process but is an iterative self-correction process. Those mechanisms have already proven to offer a remarkable increase in the robustness of collaborative AI systems by Zhang et al and Le et al [24], [25], which makes the principles of distributed intelligence valuable in relation to the complexity of the tasks in question.

*C. Adaptive Problem-Solving Strategies*

Our framework goes beyond preset processes of analysis, rather includes the versatile interaction of agents which change depending on new findings. In such cases when a category of crime can be predicted poorly by the initial predictions, the system itself begins the re-checking of specified features and variables. Such a behavior of adaptation reflects the emergent problem-solving that occurs when the framework re-organizes its method of analysis intelligently with regard to observed errors or new intelligence. This is in some sense a feature of intelligent multi-agent systems that operate in complex domains [26], [27].

*D. Collective Knowledge Construction*

By sharing outcomes on a constant basis, providing effective feedback and making improvements in coordination, our agents harness collective picture of complex crime sceneries. This decentralized but synchronized knowledge-making activity facilitates the framework to build multi-dimensional understandings about patterns of crimes at both space and time. Some kind of collective intelligence can be seen as a foundation of successful multi-agent learning systems [28], [29] and an absolute necessity in dealing with the complexity and diversity innate to urban crime data.

*E. Robustness to Data Complexity and Noise*

Crime-related data that are used in urban areas (the example of Chicago can be provided) have their problems of their own such as noise, inconsistencies, and reporting biases, as well as statistical outliers that may impair the precision of analysis. We deal with such complexities in our framework by utilizing agent perspectives in a broad sense and establishing strict cross-validation procedures. The different dimensions of data analyzed by each agent process noisy patterns but has the effect of filtering problematic noise, and reproducing valid patterns. This multi-agent redundancy and cross-checking method is one of the tried-and-true options to increase analytical reliability in non-trivial areas [30], [31] and this example proves that distributed intelligence can negate personal analytical drawbacks.

*F. "Wisdom of the Crowd" Effect within AI Agents*

By using the combined intelligence of many independently thinking agents our system is able to exploit a strong phenomenon known as a wisdom of the crowd. As has been demonstrated in many cases with heterogeneous human groups outperforming single experts due to divergent opinions and differing modes of thought [32] our multi-agent application can indeed overcome the intrinsic biases and limitation of any particular (e.g., single) LLM.

*G. Emergent Skill Chaining and Workflow Optimization*

During repetitive cycles of interactions, the agents of our framework narrow down their interaction patterns and task synchronization. Analysis Assistant, such as, learns to make more and more specific questions to the Feedback Agent, to make the process of data analysis quicker. This new business of skill chaining and streamlining work transforms minimizes redundancy and improves the performance of the whole system. This learned coordination is dynamic, the capacity to collaborate that has evolved and improved within multi-agent AI systems [33], [34] since their initial programming and it means that collaborative intelligence is capable of continuing its evolution and development so as to deliver better operational efficiency.

These findings indicate that emergent intelligence is an organic process rather than a programmed ability that just occurs naturally through the respective arrays of structure, communicative and repeated interactions of dedicated agents. Leaning on these emergent behaviors, our framework provides more advanced, robust, and versatile crime prediction and analysis functionality when faced with large and complex urban data.

## VIII. DISCUSSION

While AutoGen has primarily been used for analyzing structured crime information, recent advancements in Transformers open up exciting possibilities for expanding into visual intelligence. Notably, Vision Transformers (ViT) have shown remarkable success in identifying and analyzing crucial elements within images [36], [37], whereby such works suggest a clear path for future versions of AutoGen to incorporate visual agents. These agents could be instrumental in processing various visual data, including surveillance camera footage, crime scene details, and forensic reconstructions.

Furthermore, integrating ViT would enable a seamless multimodal system that combines crime reports with relevant images. Such a framework would significantly enhance our understanding of each situation, effectively bridging the gap between traditional data-based analysis and the rich visual information found at real crime scenes. This transformation would position AutoGen as a comprehensive, one-stop platform for tackling the complex challenges of modern crime analysis.

The results from 100 epochs reveal several valuable insights into the strengths and limitations of our AutoGen-based multi-agent system LUCID-MA for crime data analysis:

### *A. Strengths of the Framework*

- **Simulated Learning via Feedback**: Our agents showed remarkable learning-like behavior through feedback cycles, even without traditional training methods. We saw this clearly as the Crime Analysis Assistant produced increasingly refined outputs, and the Predictor Agent evolved from generic statements to actionable insights.

- **Autonomous Improvement**: As soon as it got the system up and running, it required no further human involvement. Human oversight was effectively stood in by the feedback loops, which steadily increased the level of analytical sophistication and the quality of forecasts.

- **Offline and Secure**: Finally, we ran everything locally, on our GPU Enabled DGX station and maintained privacy and security throughout, as it was still possible. If you are dealing with sensitive crime data, so there are ethical and legal issues, therefore it is important to follow this approach.

- **Modularity**: The agents are modular in design, so each one can be changed, expanded or optimized without influencing any of the other agents. However, this architecture has the ability to scale further in the future with other components such as additional components to support future scaling (Validator Agents, Law Policy Agents, or real time data processing capabilities).

### *B. Future Directions*

In a bid to facilitate future research studies and further discussion in this field, the following set of thought-provoking questions will examine the great potential and future implications emergent intelligence has to offer in the multi-agent system:

- **Qualitative Futuristic Research Question:** Is emergent intelligence in multi-agent systems evolvable into the real form of Collective Agentic Intelligence, where the distributed multi-agents are merged into a single, uniform entity that can perform high-level reasoning and decision making in artificial intelligence?

- **Mechanistic Inquiry:** How exactly can collaborative learning and emergent intelligence be observed to increase the generalization properties of the system especially in conversational and dialogue-driven environments?

### *B. Observed Limitations*

Our research revealed several areas for improvement:

- **Prompt Sensitivity**: Vulnerability in the system to instruction quality was thereby created in its reliance on prompt engineering. Our simulated learning approach helped mitigate this problem, besides a true few-shot adaptation or prompt tuning mechanism, which may make the system more robust to ambiguous instructions would be deployed, that can be seen in [35].

- **Response Redundancy**: Despite having repetition penalties in place, we observed response redundancy in the middle experimental phases. Self-observed appearance of regular patterns of output in neural degenerative situations is a phenomenon that has been well studied in the literature [36]. These patterns could be due to pathological forms of feedback or due to the disappearance of error gradients in a recurrent or iterative reasoning machine.

- **Prediction Without Validation**: The CrimePredictorAgent was able to generate insightful trend analyses, but did not have mechanisms to validate the results with actual outcomes. Verification of the forecast accuracy would make the predictive reliability over time better with the incorporation of a model retraining loop (and historical back testing module).

# IX. Conclusion

This study demonstrated the efficacy of LUCID-MA, a locally hosted AutoGen multi-agent system, in processing structured prompts for crime data analysis and subsequently generating predictive insights. Specifically, we deployed three distinct modules—a specialized analyst, a feedback critic, and a forecaster—and observed their iterative communication and performance enhancement over 100 iterations. Notably, these improvements were achieved without any additional training or direct human oversight. To quantitatively assess these changes, we developed a novel scoring methodology and visualization metrics. These tools were instrumental in structuring agent interaction, demonstrating how feedback mechanisms amplify collective intelligence over time. This highlights that AutoGen's collaborative process is capable of yielding significant novel insights, even when operating with static datasets.

In addition, this work underscores that meticulous attention to system architecture enables autonomous agents to achieve levels of comprehensiveness, depth, and adaptability comparable to traditional expert analysis. The current multi-agent paradigm is conceptually underpinned by two key areas of progress: recent advancements in Large Language Models (LLMs), particularly their enhanced efficiency, ethical alignment, and reasoning capabilities [38], and the concurrent innovations in hybrid architectures for potential vision-based agents [39]. These trends directly align with the core objectives of AutoGen: fostering agentic intelligence, specifically the capacity for iterative, context-sensitive reasoning. Moreover, collaborative learning among AI agents gives rise to emergent intelligence. This collective wisdom, stemming from shared experiences and inter-agent exchange, empowers multi-agent systems, particularly when configured as distributed Agentic AI systems, to tackle complex problems intractable for individual agents, thereby signaling a paradigm shift towards distributed artificial intelligence.

# Acknowledgment

The computational experiments for this research were performed on the NVIDIA DGX computer, configured with the open-source AutoGen framework. The authors extend their sincere gratitude to the Pattern Recognition Lab (PRLab), PIEAS AI Center (PAIC), and Center for Mathematical Sciences (CMS) at PIEAS for their invaluable support and provision of computational resources.